\documentclass[preprint,review,12pt]{elsarticle}

\usepackage{moreverb,subfigure,soul}
\usepackage{bm,color,transparent,amsmath,amssymb,amsthm,fancyhdr,mathrsfs,graphicx,setspace}
\biboptions{sort&compress}
 \usepackage[original]{pict2e}
\usepackage{diagbox,cancel} 
\usepackage[]{units}
\usepackage[framemethod=tikz]{mdframed}

\journal{Computers \& Structures}

\begin{document}

\begin{frontmatter}



\title{Multi-scale computational homogenisation to predict the long-term durability of composite structures}


\author[label1]{Z. Ullah\fnref{label2}}
\author[label1]{{\L}. Kaczmarczyk}
\author[label3,label4]{S. A. Grammatikos}
\author[label3,label4]{M. C. Evernden}
\author[label1]{C. J. Pearce}
\address[label1]{School of Engineering, Rankine Building, The University of Glasgow, Glasgow, G12 8LT, UK}
\address[label3]{BRE Centre for Innovative Construction Materials, UK}
\address[label4]{Department of Architecture and Civil Engineering, University of Bath, Bath BA2 7AY, UK}
\fntext[label2]{Correspondence to: Z. Ullah, E-mail: Zahur.Ullah@glasgow.ac.uk}

\begin{abstract}
A coupled hygro-thermo-mechanical computational model is proposed for fibre reinforced polymers, formulated within the framework of Computational Homogenisation (CH). At each macrostructure Gauss point, constitutive matrices for thermal, moisture transport and mechanical responses are calculated from CH of the underlying representative volume element (RVE). A degradation model, developed from experimental data relating evolution of mechanical properties over time for a given exposure temperature and moisture concentration is also developed and incorporated in the proposed computational model. A unified approach is used to impose the RVE boundary conditions, which allows convenient switching between linear Dirichlet, uniform Neumann and periodic boundary conditions. A plain weave textile composite RVE consisting of yarns embedded in a matrix is considered in this case. Matrix and yarns are considered as isotropic and transversely isotropic materials respectively. Furthermore, the computational framework utilises hierarchic basis functions and designed to take advantage of distributed memory high performance computing.

\end{abstract}


\end{frontmatter}


\section{Introduction}\label{sec_intro}

Fibre reinforced polymer (FRP) composites have exceptional mechanical and chemical properties, including light weight, high specific strength, fatigue and corrosion resistance, low thermal expansion and high dimensional stability. They are commonly used in engineering application including aerospace, ships, offshore platforms, automotive industry, prosthetics and civil structures \cite{Tong2002, Mouritz1999}. Textile or woven composites is a class of FRP composites, in which interlaced fibres are used as reinforcement, which provides full flexibility of design and functionality due to the mature textile manufacturing industry \cite{Long2005}. A detailed review, explaining the design and fabrication of textile preforms including weaving, knitting, stitching and braiding with their potential advantages and limitations is given in \cite{Kamiya2000}.  As compared to the standard laminated composites, textile composites have better damage and impact resistance, better through-thickness properties and reduced manufacturing cost. However, waviness of the yarns in the textile composites reduces the tensile and compressive strengths \cite{Miravete1999}.  

Due to their complicated and heterogeneous microstructure, Computational Homogenisation (CH) provides an accurate modelling framework to simulate the behaviour of FRP composites and determine the macro-scale homogenised (or effective) properties, including mechanical stiffness, thermal conductivity and moisture diffusivity, based on the physics of an underlying, microscopically heterogeneous, representative volume element (RVE)  \cite{NematNaseer1993, Geers2010, Ullah_ACME2014, Ullah_ACME2015, Sankar1997}. The homogenised properties calculated from the multi-scale CH are subsequently used in the numerical analysis of the macro-level structure. A variety of analytical and numerical homogenisation schemes have been developed for textile based FRP composites, which are normally based on the existence of an RVE and focus attention on the mechanical behaviour.  Analytical methods are quick and easy to use but generally give poor estimates of the homogenised properties and are normally based on oversimplified assumptions of the microstructure and states of stress and strain. In the literature, some of the analytical homogenisation schemes, with their potential applications and limitations highlighted, are given in \cite{ yang1986, Naik_Nasa_1994, Sankar1997, Brian_NASA_1997, Hallal2013}. Numerical techniques, on the other hand, can accurately estimate the homogenised properties by capturing accurately the intricate micro-structure exactly but are computationally expensive.  Examples of numerical homogenisation schemes applied to FRP composites can be found in \cite{Yoshino1982, Dasgupta1994, Daggumati2010, Bobzin2010, Gager2012, Jacques2014}.  A review article, summarising some of the analytical and numerical homogenisation techniques for the mechanical properties of the textile composite is given in \cite{Tan1997}.  

During their service lives, FRP structures can be exposed to harsh hygro-thermal environmental conditions in addition to mechanical loading, which can lead to matrix plasticisation, hydrolysis and degradation of fibres/matrix interfaces \cite{Tang2005, Tang_PhD_2001, Bailakanavar2014}. In the long-term, these processes significantly reduce the mechanical performance of these structures. Therefore, understanding heat and moisture transport mechanisms and their effect on the mechanical performance are fundamental for assessing the long-term use of FRP structures. Effective diffusivities of FRP composites with impermeable fibres were studied in \cite{whitcomb2002} within the context of FEM, considering the variation in fibre volume within square and hexagonal unit cells. In \cite{Yang_2010}, effective moisture diffusivity as a function of temperature and fibre volume fraction were investigated for FRP composites with permeable fibres using a unit cell approach. For textile composites, moisture transport was studied in \cite{Tang_PhD_2001, Tang2005} as a function of variation in tow architectural parameters, e.g. tow waviness, tow cross-section shape and wave pattern. Transient moisture transport in multilayer textile composites was investigated in \cite{Pasupuleti2011}. An analogy between thermal and moisture transport analysis was used in \cite{Laurenzi2008} to study the moisture diffusion and corresponding weight gain for carbon braided composites. In \cite{Bailakanavar_PhD_2013, Bailakanavar2014}, a multi-scale CH framework based on the hygro-mechanical analysis was proposed while using a two-dimensional RVE with randomly distributed fibres in $0^o$ and $90^o$ directions. A two way coupling was considered between the mechanical and moisture transport analysis and a model reduction scheme was used to reduce the computational cost. For the composite material, deformation dependent diffusion at finite strains was considered in \cite{Klepach2014}. Masonry wall reinforced with FRP reinforcement was studied in \cite{Khoshbakht2010} while considering hygro-thermo-mechanical analysis. A recent review article \cite{Ray2015}, explains different degradation mechanism for FRP composites in connection with different environmental conditions. 

In this paper, a coupled hygro-thermo-mechanical computational framework based on the multiscale CH is proposed for FRP composites. At each integration point, an RVE consisting of single plain weave textile composite is considered, consisting of yarns embedded in the matrix. Elliptical cross sections and cubic spline paths are used to model the geometry of these yarns. Separate RVEs are considered for the heat transfer, moisture transport and mechanical CH. One-way coupling is considered in this case, i.e. mechanical analysis is assumed to be dependent on both moisture transport and thermal analyses but any influence on the moisture or thermal behaviour due to the mechanical behaviour is ignored. A degradation model, developed from experimental data relating evolution of mechanical stiffness over time for given exposure temperatures and moisture concentration was also developed and incorporated in the proposed computational framework. A unified approach is used to impose the RVE boundary conditions, which allows convenient switching between the different RVEs boundary conditions (linear Dirichlet, uniform Neumann and periodic) \cite{Lukasz2008}. For a given size of RVE the periodic boundary conditions gives a better estimation of the homogenised material properties as compared to linear Dirichlet and uniform Neumann boundary conditions, which give an upper and lower limit \cite{Kouznetsova_PhD_2002, Lukasz2008, Miehe_2002} respectively. 

The developed computational framework utilises the flexibility of hierarchic basis functions \cite{Ainsworth2003}, which permits the use of arbitrary order of approximation, thereby improving accuracy for relatively coarse meshes. For the thermal and moisture transport analyses both matrix and yarns are assumed as isotropic materials, while for the mechanical analysis, the yarns are considered as transversely isotropic materials. The required principal directions of the yarns for the transversely isotropic material model are calculated from potential flow analysis along these yarns. Furthermore, distributed memory high performance computing is used to reduce the computational cost associated with the current multi-scale and multi-physics computational framework.

This paper is organised as follows. The multi-scale CH framework and corresponding implementation of the RVE boundary conditions are described in \S \ref{sec_compHomo}. Transient heat and moisture transport analyses along with their FE implementation are discussed in \S \ref{sec_transFieldProblem}. The derivation of the degradation model from the experimental data is next explained in \S \ref{sec_degradation}. The overall multi-scale and multi-physics computational framework is described in detail in \S \ref{sec_comp_framework}. Computation of yarns directions are explained in \S \ref{sec_fibre_directions}. A three-dimensional numerical example and concluding remarks are given in \S \ref{sec_num_exp} and \S \ref{sec_conc_remarks} respectively.

\section{Multi-scale computational homogenisation}\label{sec_compHomo}
\begin{figure}[h!]
\begin{centering}
\includegraphics[scale=0.7]{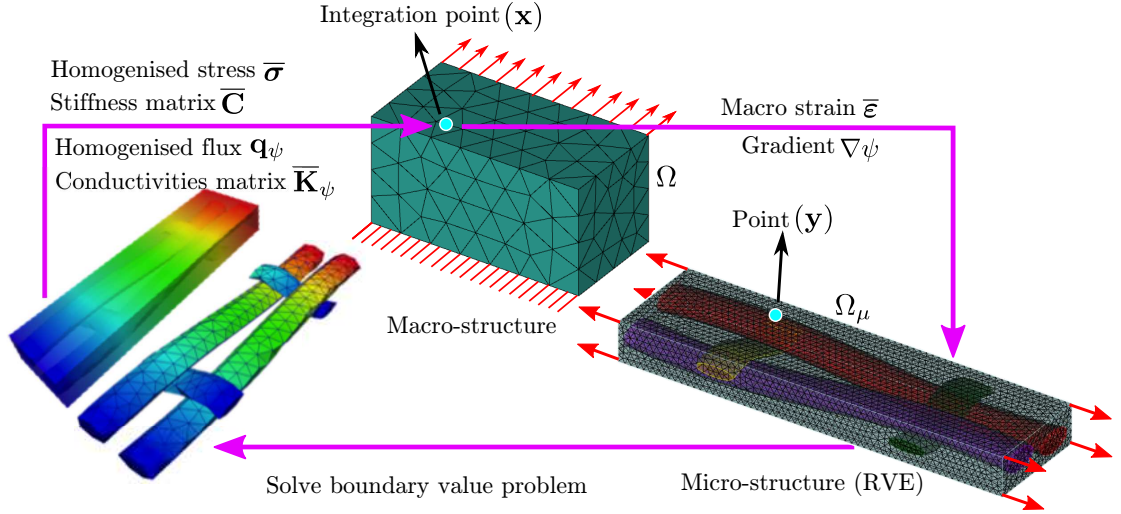}
\caption{Multi-scale computational homogenisation} \label{Fig_CH}
\end{centering}
\end{figure}

In multi-scale CH, a heterogeneous RVE is associated with each Gauss point of the macro-homogeneous structure. Multi-scale CH gives us directly the macro-level constitutive relation, allows us to incorporate large deformation and rotation on both micro- and macro-level and both physical and geometric evolution can be included on both micro- and macro-level \cite{Kouznetsova_PhD_2002}. The multi-scale CH procedure and corresponding implementation of RVE boundary conditions is described initially for the mechanical case, which is subsequently extended to corresponding thermal and moisture transport cases. The first order multi-scale CH is used in the paper, the basic principle of which is shown in Figure \ref{Fig_CH}, where $\Omega\subset\mathbb{R}^{3}$ and $\Omega_{\mu}\subset\mathbb{R}^{3}$ are macro and micro domains respectively. 
Macro-strain $\overline{\boldsymbol{\varepsilon}}=\left[\begin{array}{cccccc}\overline{\varepsilon}_{11} & \overline{\varepsilon}_{22} & \overline{\varepsilon}_{33} & 2\overline{\varepsilon}_{12} & 2\overline{\varepsilon}_{23} & 2\overline{\varepsilon}_{31}\end{array}\right]^{T}$ is first calculated at each Gauss point $\mathbf{x}=\left[\begin{array}{ccc}x_{1} & x_{2} & x_{3}\end{array}\right]^T$ of the macro-structure, which is then used to formulate the boundary value problem on the micro-level. After solution of the micro-level boundary value problem, homogenised stress $\overline{\boldsymbol{\sigma}}=\left[\begin{array}{cccccc}\overline{\sigma}_{11} & \overline{\sigma}_{22} & \overline{\sigma}_{33} & \overline{\sigma}_{12} & \overline{\sigma}_{23} & \overline{\sigma}_{31}\end{array}\right]^{T} $ and stiffness matrix $\overline{\mathbf{C}}$ are calculated. Separation of scales is assumed in the first-order CH, i.e. the micro length scale is considered to be very small compared to the macro length scale and the macro-strain field attributed to each RVE is assumed to be uniform. Therefore, first-order CH is not suitable for problems with large strain gradient (but can be used for problems subjected to large strain) and cannot be used to take into account micro-level geometric size effects \cite{Kouznetsova_PhD_2002, Geers2010}. On the micro-level at any point $\mathbf{y}=\left[\begin{array}{ccc}y_{1} & y_2 & y_3\end{array}\right]^{T}$ the displacement field is written as \cite{Peric2007, Peric2011, Xiaoyi2015}
\begin{equation}\label{micro_disp}
\mathbf{u}_{\mu}\left(\mathbf{y}\right)=\overline{\boldsymbol{\varepsilon}}\left(\mathbf{x}\right)\mathbf{y}+\widetilde{\mathbf{u}}_{\mu}\left(\mathbf{y}\right),
\end{equation}
where $\overline{\boldsymbol{\varepsilon}}\mathbf{y}$ is a linear displacement field and $\widetilde{\mathbf{u}}_{\mu}$ is a displacement fluctuation. The micro-strain associated with point $\mathbf{y}$ is written as
\begin{equation}\label{micro_strain}
\boldsymbol{\varepsilon}_{\mu}\left(\mathbf{y}\right)=\nabla^{s}\mathbf{u}_{\mu}=\overline{\boldsymbol{\varepsilon}}\left(\mathbf{x}\right)+\tilde{\boldsymbol{\varepsilon}}\left(\mathbf{y}\right),
\end{equation}
where $\tilde{\boldsymbol{\varepsilon}}\left(\mathbf{y}\right)=\nabla^{s}\tilde{\mathbf{u}}_{\mu}$ is the strain fluctuation at the micro-level and $\nabla^{s}$ is the symmetric gradient operator. Furthermore, volume average of the micro-strain is equivalent to the macro-strain:
\begin{equation}\label{micromacro_strain}
\overline{\boldsymbol{\varepsilon}}\left(\mathbf{x}\right)=\frac{1}{V}\int_{\Omega_{\mu}}\boldsymbol{\varepsilon}_{\mu}\left(\mathbf{y}\right)d\Omega_{\mu}=\overline{\boldsymbol{\varepsilon}}\left(\mathbf{x}\right)+\frac{1}{V}\int_{\Omega_{\mu}}\tilde{\boldsymbol{\varepsilon}}_{\mu}\left(\mathbf{y}\right)d\Omega_{\mu},
 \end{equation}
where $V$ is the volume of the RVE. It is clear from Equation (\ref{micromacro_strain}) that the volume average of the strain fluctuation is zero, i.e. 
\begin{equation}
\frac{1}{V}\int_{\Omega_{\mu}}\tilde{\boldsymbol{\varepsilon}}_{\mu}\left(\mathbf{y}\right)d\Omega_{\mu}=0.
\end{equation}
The micro-equilibrium state in  the absence of body force is written as
\begin{equation}
\textrm{div}\left(\boldsymbol{\sigma}_{\mu}\right)=\nabla\cdot\boldsymbol{\sigma}_{\mu}=\mathbf{0},
\end{equation}
or in the variational form 
\begin{equation}\label{RVE_Eq_var}
\int_{\Omega_{\mu}}\boldsymbol{\sigma}_{\mu}\left(\mathbf{y}\right):\nabla^{s}\boldsymbol{\eta}_{\mu}d\Omega_{\mu}-\int_{\partial\Omega_{\mu}}\mathbf{t}\left(\mathbf{y}\right)\cdot\boldsymbol{\eta}_{\mu}\partial\Omega_{\mu}=\mathbf{0},
\end{equation}
where $\mathbf{t}$ is an external applied traction and $\boldsymbol{\eta}_{\mu}$ is a virtual displacement.  No constitutive assumption is made for the macro-level problem, and the macro-stress $\overline{\boldsymbol{\sigma}}$ is determined by volume averaging of the micro-stress $\overline{\boldsymbol{\sigma}}_{\mu}$, i.e.  
\begin{equation}\label{macro_stress}
\overline{\boldsymbol{\sigma}}\left(\mathbf{x}\right)=\frac{1}{V}\int_{\Omega_{\mu}}\boldsymbol{\sigma}_{\mu}\left(\mathbf{y}\right)d\Omega_{\mu}.
\end{equation}
To determine consistent RVE boundary conditions Hill-Mandel principle is used, which equates the work done at the micro and macro levels, i.e.
\begin{equation}\label{Hill_prin}
\overline{\boldsymbol{\varepsilon}}:\overline{\boldsymbol{\sigma}}=\frac{1}{V}\int_{\Omega_{\mu}}\boldsymbol{\varepsilon}_{\mu}:\boldsymbol{\sigma}_{\mu}d\Omega_{\mu},
\end{equation}
After combining Equations (\ref{Hill_prin}), (\ref{macro_stress}), (\ref{RVE_Eq_var}) and (\ref{micro_disp}), we obtain 
\begin{equation}
\int_{\partial\Omega_{\mu}}\mathbf{t}\left(\mathbf{y}\right)\cdot\boldsymbol{\eta}\partial\Omega_{\mu}=0,
\end{equation}
i.e. the virtual work associated with the traction $\mathbf{t}$ vanishes and the variational form of the equilibrium Equation  (\ref{RVE_Eq_var}) reduce to 
\begin{equation}\label{RVE_Eq_var_final}
\int_{\Omega_{\mu}}\boldsymbol{\sigma}_{\mu}\left(\mathbf{y}\right):\nabla^{s}\boldsymbol{\eta}_{\mu}d\Omega_{\mu}=0.
\end{equation}
Thus, on the RVE level, the problem reduces to the calculation of the displacement fluctuation $\widetilde{\mathbf{u}}_{\mu}$ for a given macro-strain $\overline{\boldsymbol{\varepsilon}}$.

In this work, we consider three types of RVE boundary conditions, which can be shown to satisfy the Hill-Mandel principle \cite{Miehe_2002,Lukasz2008}
\begin{enumerate}
\item Linear boundary displacement  (Dirichlet): In this case, the displacement fluctuation $\widetilde{\mathbf{u}}_{\mu}$ are assumed as zero over the boundary of the RVE, which leads to fully prescribed displacements on the RVE boundary
\begin{equation}
\mathbf{u}_{\mu}\left(\mathbf{y}\right)=\overline{\boldsymbol{\varepsilon}}\left(\mathbf{x}\right)\mathbf{y},\qquad\forall\mathbf{y}\in\partial\Omega_{\mu}.
\end{equation}

\item Periodic boundary conditions: In this case it is assumed that displacement fluctuation $\widetilde{\mathbf{u}}_{\mu}$ is periodic and traction is anti-periodic, i.e. 
\begin{equation}
\left.\begin{array}{c}
\tilde{\mathbf{u}}_{\mu}\left(\mathbf{y}^{+}\right)=\tilde{\mathbf{u}}_{\mu}\left(\mathbf{y}^{-}\right)\\
\mathbf{t}\left(\mathbf{y}^{+}\right)=-\mathbf{t}\left(\mathbf{y}^{-}\right)
\end{array}\right\} \qquad\forall\,\,\textrm{pairs}\left\{ \begin{array}{cc}
\mathbf{y}^{+}, & \mathbf{y}^{-}\end{array}\right\} ,
\end{equation}
where $\mathbf{y}^{+}$ and $\mathbf{y}^{-}$ are a pair of points on the opposite boundary of the RVE, i.e. $\partial\Omega_{\mu}^{+}$ and $\partial\Omega_{\mu}^{-}$ respectively. 

\item Uniform traction  (Neumann) boundary conditions: In this case the traction on the boundary of the RVE is written in term of the macro-stress  $\overline{\boldsymbol{\sigma}}$, i.e. 
\begin{equation}
\mathbf{t}=\overline{\boldsymbol{\sigma}}\cdot\mathbf{n}, 
\end{equation}
where $\mathbf{n}$ is the outward normal of the RVE boundary. 
\end{enumerate}
The imposition of these different type of RVE boundary conditions leads to different responses of the RVE. The linear displacement boundary condition gives the stiffest response followed by periodic, while the uniform traction boundary condition imposes the least kinematic constraint. In this paper, RVE boundary conditions are prescribed using the procedure given in \cite{Lukasz2008} with extension to three-dimensions. The final discretised system of equations in the case of the RVE is written as
\begin{equation}\label{eq_RVE_disc_Eqs}
\left[\begin{array}{cc}
\mathbf{K} & \mathbf{C}^{T}\\
\mathbf{C} & \mathbf{0}
\end{array}\right]\left\{ \begin{array}{c}
\mathbf{u}\\
\boldsymbol{\lambda}
\end{array}\right\} =\left\{ \begin{array}{c}
\mathbf{0}\\
\mathbf{D}\overline{\boldsymbol{\varepsilon}}
\end{array}\right\}, 
\end{equation}
where $\mathbf{K}$ and $\mathbf{u}$ are the standard FE stiffness matrix and displacement vector and $\boldsymbol{\lambda}$ is the unknown vector of Lagrange multipliers required to impose the RVE boundary conditions. Matrices $\mathbf{C}$ and $\mathbf{D}$ are given as 
\begin{equation}\label{eq_mat_C_and_D}
\mathbf{C}=\int_{\partial\Omega_{\mu}}\mathbf{HN}^{T}\mathbf{N}d\partial\Omega_{\mu}, \qquad \mathbf{D}=\int_{\partial\Omega_{\mu}}\mathbf{HN}^{T}\mathbf{X}d\partial\Omega_{\mu}.
\end{equation}
In Equation (\ref{eq_mat_C_and_D}), $\mathbf{N}$ is a matrix of shape functions and $\mathbf{X}$ is a matrix of spatial coordinates, evaluated at Gauss points during numerical integration of the surface integrals in Equation (\ref{eq_mat_C_and_D}) and is given as
\begin{equation}\label{eq_X_mech}
\mathbf{X}=\frac{1}{2}\left[\begin{array}{cccccc}
2y_1 & 0 & 0 & y_2 & y_3 & 0\\
0 & 2y_2 & 0 & y_1 & 0 & y_3\\
0 & 0 & 2y_3 & 0 & y_1 & y_2
\end{array}\right].
\end{equation}
Furthermore, $\mathbf{H}$ is a matrix that is specific to the type of boundary conditions used, each row of which represents an admissible distribution of nodal traction forces on the RVE boundary \cite{Lukasz2008}. The specific choice of $\mathbf{H}$ in case of linear displacement, periodic and uniform traction boundary conditions can be found in \cite{Lukasz2008, Xiaoyi2015} and is not repeated here. Method of static condensation and projection matrices are alternative procedures to solve system of equations (\ref{eq_RVE_disc_Eqs}) \cite{Lukasz2008}. 

To determine the homogenised stress, the right hand side of Equation (\ref{Hill_prin}) is written in term of surface quantities, i.e.
\begin{equation}\label{eq_Hill_prin_surf1}
\overline{\boldsymbol{\varepsilon}}:\overline{\boldsymbol{\sigma}}=\frac{1}{V}\int_{\partial\Omega_{\mu}}\mathbf{t}\cdot\mathbf{u}_{\mu}d\partial\Omega_{\mu}. 
\end{equation}
Furthermore, the work done by the traction on displacements is equal to the work of the generalised tractions (Lagrange multiplers) on the generalised displacements (strains):
\begin{equation}\label{eq_Hill_prin_surf2}
\mathbf{u}^T\mathbf{t}={\left(\mathbf{D}\overline{\boldsymbol{\varepsilon}}\right)}^T\boldsymbol{\lambda}.
\end{equation}
With reference to Equations (\ref{eq_Hill_prin_surf1}) and (\ref{eq_Hill_prin_surf2}), the homogenised stress can be determined as:
\begin{equation}\label{eq_homo_stress}
\overline{\boldsymbol{\sigma}}=\frac{1}{V}\mathbf{D}^T\boldsymbol{\lambda}
\end{equation}
For the macro-level finite element analysis, material stiffness matrix $\mathbf{C}$ need to be determined at each Gauss point. The relationship between macro-stress and strain is written as 
\begin{equation}
\overline{\boldsymbol{\sigma}}=\overline{\mathbf{C}}\overline{\boldsymbol{\varepsilon}}.
\end{equation}
To compute the homogenised stiffness matrix $\overline{\mathbf{C}}$, six linear system of equations for the RVE need to be solved for a given set of unit strain vectors. This will give a set of homogenised stresses, i.e.
\begin{equation}\label{Eq_C_stress}
\overline{\mathbf{C}}=\left[\begin{array}{cccccc}
\overline{\boldsymbol{\sigma}}^{1} & \overline{\boldsymbol{\sigma}}^{2} & \overline{\boldsymbol{\sigma}}^{3} & \overline{\boldsymbol{\sigma}}^{4} & \overline{\boldsymbol{\sigma}}^{5} & \overline{\boldsymbol{\sigma}}^{6}\end{array}\right], 
\end{equation}
where for example: 
\begin{equation}
\begin{array}{cc}
\overline{\boldsymbol{\sigma}}^{1}: & \textrm{for}\,\,\,\,\overline{\boldsymbol{\varepsilon}}=\left[\begin{array}{cccccc}
1 & 0 & 0 & 0 & 0 & 0\end{array}\right]^{T}\\
\overline{\boldsymbol{\sigma}}^{4}: & \textrm{for}\,\,\,\,\overline{\boldsymbol{\varepsilon}}=\left[\begin{array}{cccccc}
0 & 0 & 0 & 1 & 0 & 0\end{array}\right]^{T}\\
\end{array}.
\end{equation}
In each of the six cases, only the right-hand side of the system of Equations (\ref{eq_RVE_disc_Eqs}) changes, which is solved very efficiently as the left-hand side matrix is factorised only once.

Computational homogenisation for thermal and moisture transport cases are very similar and is presented as a single case in the following. The complete derivation is not repeated due to its similarity with the mechanical counterpart but only the key differences are highlighted. The gradient of macro field $\nabla\psi=\left[\begin{array}{ccc}
\frac{\partial\psi}{\partial x_{1}} & \frac{\partial\psi}{\partial x_{2}} & \frac{\partial\psi}{\partial x_{3}}\end{array}\right]^{T}$ is used to formulate the boundary value problem on the micro-level, where $\psi=T, c$ is a scalar field (temperature and moisture concentration respectively). The homogenised flux $\mathbf{q}_{\psi}=\left[\begin{array}{ccc}
q_{\psi}^{x_{1}} & q_{\psi}^{x_{2}} & q_{\psi}^{x_{3}}\end{array}\right]^{T}$ and corresponding conductivity matrix $\overline{\mathbf{K}}_{\psi}$ are next obtained after solution of RVE's boundary value problem. The complete homogenisation process for this case is also shown in Figure \ref{Fig_CH}. Similar to Equation (\ref{eq_RVE_disc_Eqs}), the final discretised system of equations in this case is written as: 
\begin{equation}\label{eq_RVE_disc_Eqs_field}
\left[\begin{array}{cc}
\mathbf{K}_{\psi} & \mathbf{C}_{\psi}^{T}\\
\mathbf{C_{\psi}} & \mathbf{0}
\end{array}\right]\left\{ \begin{array}{c}
\boldsymbol{\psi}_{\mu}\\
\boldsymbol{\lambda}\psi
\end{array}\right\} =\left\{ \begin{array}{c}
\mathbf{0}\\
\mathbf{D}\nabla\psi
\end{array}\right\},
\end{equation}         
where $\mathbf{K}_{\psi}$ and $\boldsymbol{\psi}_{\mu}$ are the standard FE conductivity matrix and vector of field values (temperature or moisture concentration) respectively. $\boldsymbol{\lambda}\psi$ is a vector of Lagrange multipliers, requires to impose the RVE boundary conditions. In contrast to the three Lagrange multipliers per node for the mechanical case, only one Lagrange multiplier per node is required for temperature and moisture transport RVEs. Equations for both $\mathbf{C}_\psi$ and $\mathbf{D}$ remains the same as their mechanical counterpart in Equation (\ref{eq_mat_C_and_D}), albeit with reduced dimensions due to there being only one degree of freedom per node. The matrix for special coordinates, Equation (\ref{eq_X_mech}) is modified as    
\begin{equation}\label{eq_X_psi}
\mathbf{X}=\left[\begin{array}{ccc}
y_1 & 0 & 0 \\
0 & y_2 & 0 \\
0 & 0 & y_3
\end{array}\right].
\end{equation}
Furthermore, with reference to Equation (\ref{eq_homo_stress}), the homogenised flux is written as:  
\begin{equation}\label{eq_homo_flux}
\mathbf{q}_\psi=\frac{1}{V}\mathbf{D}^T\boldsymbol{\lambda}_\psi. 
\end{equation}
The homogenised flux $\mathbf{q}_\psi$ and gradient of field values $\nabla\psi$ are related with the following equation:   
\begin{equation}
\mathbf{q}_\psi=\overline{\mathbf{K}}_\psi\nabla\psi. 
\end{equation}

\section{Macro-level transient field problems}\label{sec_transFieldProblem}
FRP structures on the macro-level are generally exposed to hygro-thermal environmental conditions and require a detailed heat and moisture transport analysis. In this paper, conduction and diffusion models are considered for the heat and moisture transport analysis respectively, both of which can be represented by the following governing equation (conservation of energy for heat transfer and conservation of mass for moisture transport)
\begin{equation}\label{Eq_field}
\rho c_{p}\frac{\partial\psi}{\partial t}=\overline{K}_{\psi}^{x_1}\frac{\partial^2\psi}{\partial {x_1}^{2}}+\overline{K}_{\psi}^{x_2}\frac{\partial^2\psi}{\partial {x_2}^{2}}+\overline{K}_{\psi}^{x_3}\frac{\partial^2\psi}{\partial {x_3}^{2}},
\end{equation}
where t is time, $\rho$ and $c_p$ are macro-level density and specific heat capacity respectively. $\overline{K}_{\psi}^{x_1}$,  $\overline{K}_{\psi}^{x_2}$ and  $\overline{K}_{\psi}^{x_3}$ are thermal or moisture conductivities in $x_1$, $x_2$ and $x_3$ directions respectively, and are assumed to be independent of temperature $T$ or moisture concentration $c$. In Equation (\ref{Eq_field}), it is assumed that heat and moisture transport are governed by Fourier's law and Fick's law respectively. Fourier's law for the heat conduction is expressed as 
\begin{equation}\label{Eq_FourierLaw}
\mathbf{q}_T=-\overline{\mathbf{K}}_T\nabla T,
\end{equation}
where $\mathbf{q}_T=\left[\begin{array}{ccc}q_T^{x_1} & q_T^{x_2} & q_T^{x_3}\end{array}\right]^{T}$ is a vector of heat fluxes and $\overline{\mathbf{K}}_T$ is matrix of thermal conductivities and is written as
\begin{equation}
\overline{\mathbf{K}}_{T}=\left[\begin{array}{ccc}
\overline{K}_{T}^{x_1} & 0 & 0\\
0 & \overline{K}_{T}^{x_2} & 0\\
0 & 0 & \overline{K}_{T}^{x_3}
\end{array}\right].
\end{equation}
Similarly, Fick's law is written as 
\begin{equation}\label{Eq_FkicksLaw}
\mathbf{q}_c=-\overline{\mathbf{D}}_c\nabla c,
\end{equation}
where $\mathbf{q}_c=\left[\begin{array}{ccc}q_c^{x_1} & q_c^{x_2} & q_c^{x_3}\end{array}\right]^{T}$ is a vector of moisture fluxes and $\overline{\mathbf{D}}_c$ is matrix of moisture diffusivities and is written as
\begin{equation}
\overline{\mathbf{D}}_{c}=\left[\begin{array}{ccc}
\overline{D}_{c}^{x_1} & 0 & 0\\
0 & \overline{D}_{c}^{x_2} & 0\\
0 & 0 & \overline{D}_{c}^{x_3}
\end{array}\right],
\end{equation}
where $\overline{D}_{c}^{x_1}$,  $\overline{D}_{c}^{x_2}$ and  $\overline{D}_{c}^{x_3}$ are moisture diffusivities in $x_1$, $x_2$ and $x_3$ directions. Using Equation (\ref{Eq_field}), the relationship between diffusivity and conductivity is written as 
\begin{equation}
\overline{D}_\psi=\frac{\overline{K}_\psi}{\rho c_p}.
\end{equation}

Equation (\ref{Eq_field}) is discretised in the standard way:
\begin{equation}\label{Eq_T_disc}
\psi=\boldsymbol{\mathcal{N}}\boldsymbol{\psi}. 
\end{equation}
with $\boldsymbol{\mathcal{N}}=\left[\begin{array}{cccc}N_{1} & N_{2} & \cdots & N_{n}\end{array}\right]$ and $\boldsymbol{\psi}=\left[\begin{array}{cccc}\psi_{1} & \psi_{2} & \cdots & \psi_{n}\end{array}\right]^T$ the vectors of shape functions and nodal degrees of freedom. Thus, the discretised form Equation (\ref{Eq_field}) is written as \cite{Lewis_Book2005}
\begin{equation}\label{Eq_field_disc}
\boldsymbol{\mathcal{C}}\frac{\partial \boldsymbol{\psi}}{\partial t} + \boldsymbol{\mathcal{K}}_\psi \boldsymbol{\psi} = \mathbf{f},
\end{equation}
where 
\begin{equation}
\boldsymbol{\mathcal{C}} = \int_\Omega \rho c_p\boldsymbol{\mathcal{N}}^T \boldsymbol{\mathcal{N}} d \Omega, 
\end{equation}
\begin{equation}
\boldsymbol{\mathcal{K}}_\psi = \int_\Omega \mathbf{B}^T  \overline{\mathbf{K}}_\psi \mathbf{B} d \Omega,
\end{equation}
and 
\begin{equation}
\mathbf{f} = \int_{{\partial\Omega}_q} q_s \boldsymbol{\mathcal{N}}^T d \partial\Omega_q,
\end{equation}
where $q_s=\mathbf{q}_\psi\mathbf{n}$ with $\mathbf{n}=\left[\begin{array}{ccc}n_{x_1} & n_{x_2} & n_{x_3}\end{array} \right]^T$, the vector of normal direction cosines. Finally, $\mathbf{B} $ is the matrix of shape functions derivatives and $\partial\Omega_q$ is the boundary with specified fluxes.

\section{Degradation model}\label{sec_degradation}
A fully generalised degradation model has been developed based on experimental data for the mechanical properties of FRP composites subjected to different hygro-thermal environmental conditions. The experimental data provided by our project partner at the University of Bath, UK, involves accelerated ageing, i.e. immersing of FRP composites samples in hot distilled water with exposure temperatures of $25^oC$, $40^oC$, $60^oC$ and $80^oC$, where different mechanical parameters (tensile strength, shear strength, young modulus and shear modulus) are recorded after 0, 28, 56 and 112 days. The proposed model can be used to predict the mechanical properties of FRP composites with given histories of environmental exposure temperature and moisture concentration. The FRP composite used for the experimental studies is pultruded E-glass fibre reinforced (isophthalic polyester) polymer. 

\begin{figure}[h!]
\begin{centering}
\includegraphics[scale=0.7]{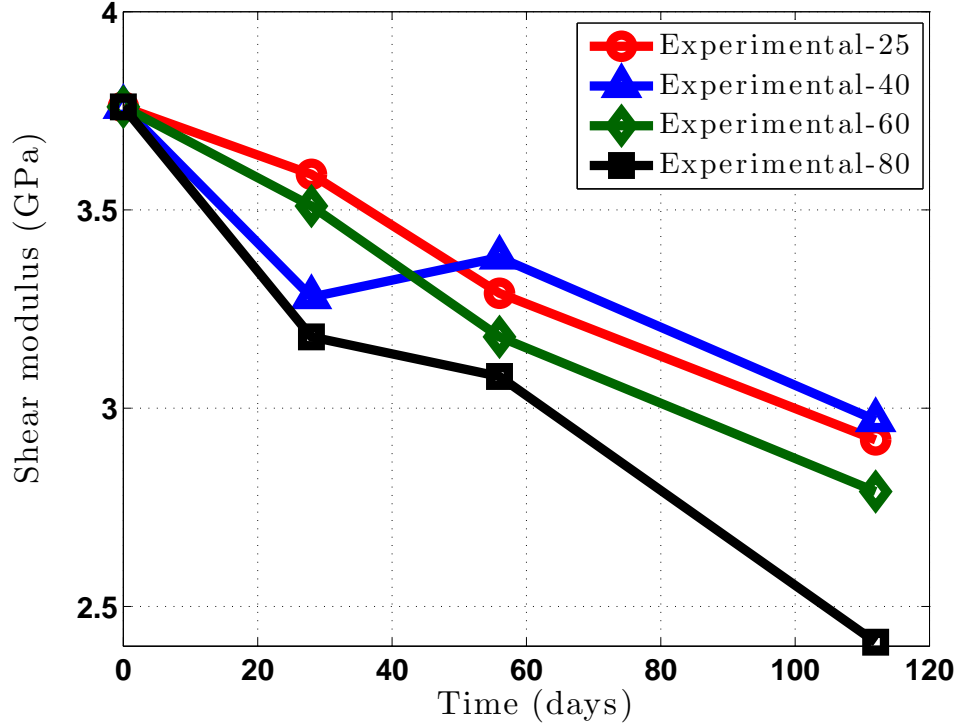}
\caption{Experimental data for the degradation of shear modulus over time for different temperatures} \label{Fig_GExp}
\end{centering}
\end{figure}

\begin{figure}[h!]
\begin{centering}
\includegraphics[scale=0.7]{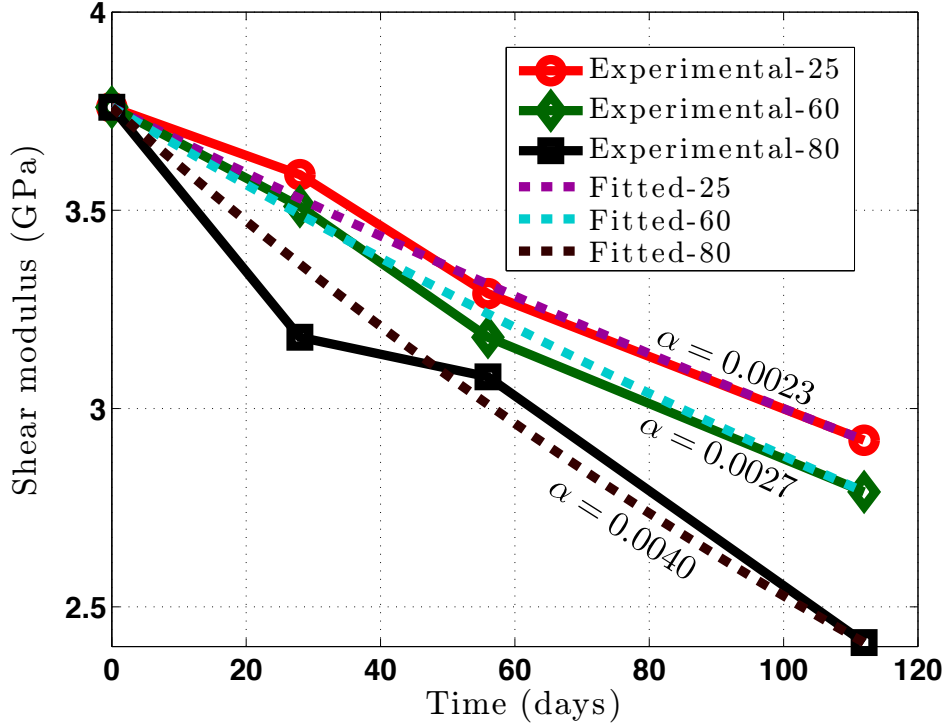}
\caption{Degradation of shear modulus over time for different temperatures (experimental and fitted exponential curves)} \label{Fig_GExp_Fit}
\end{centering}
\end{figure}

\begin{figure}[h!]
\begin{centering}
\includegraphics[scale=0.7]{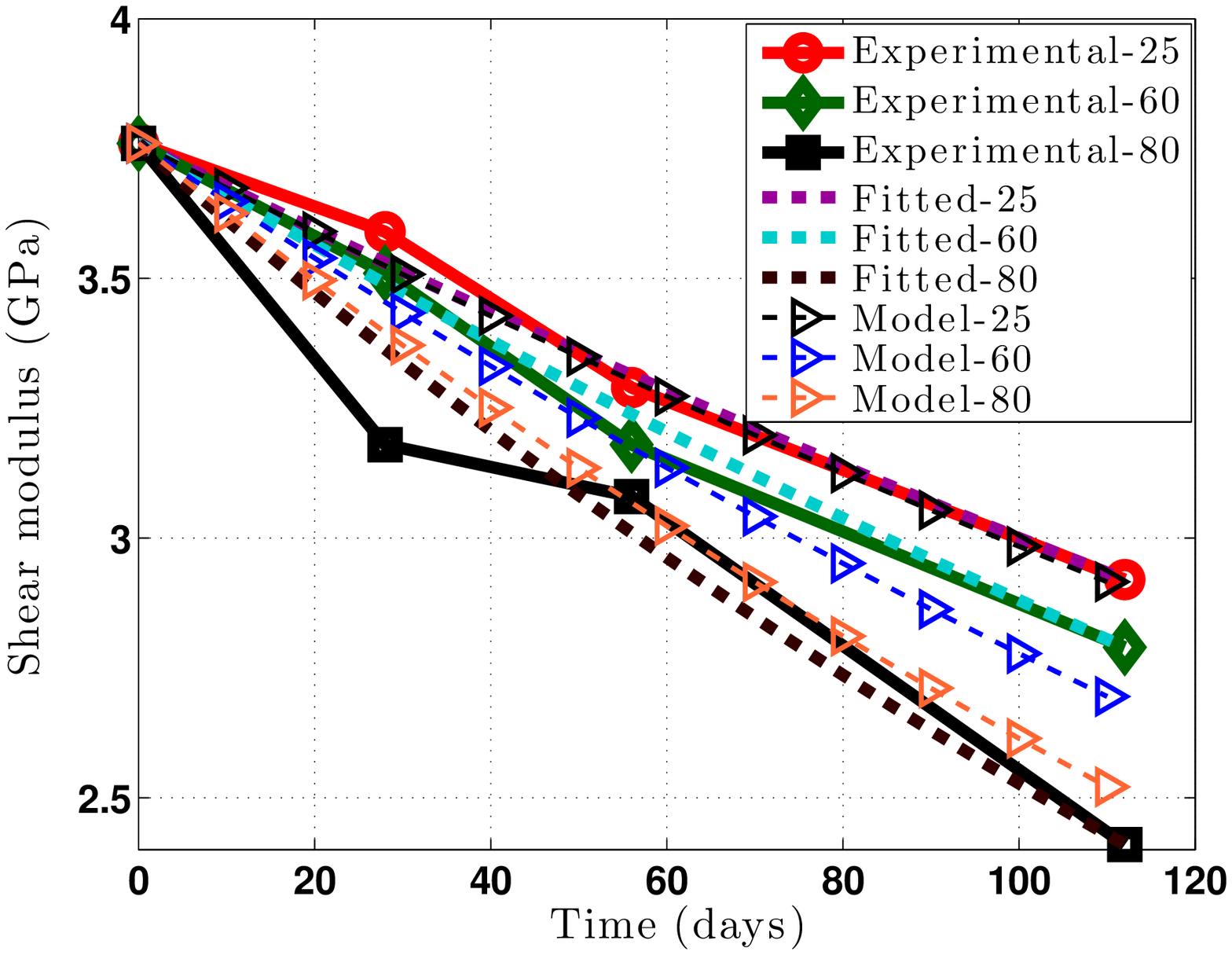}
\caption{Comparison of model predictions with experimental and fitted values of shear modulus} \label{Fig_GExp_Fit_Model}
\end{centering}
\end{figure}

The experimental data for the degradation of shear modulus with ageing for different exposure temperature is shown in Figure \ref{Fig_GExp}, which shows a clear decreasing trend for the exposure temperatures of $25^oC$, $60^oC$ and $80^oC$. The degradation data for the exposure temperature $40^oC$ is not consistent with the other three and is ignored in the subsequent analysis. An exponential trend of the form
\begin{equation}\label{Eq_ExpTrend}
G\mid_T\left(t\right)=G_oe^{-\alpha t},
\end{equation}
is assumed, where $T$ is exposure temperature in $^oC$, $G\mid_T$ is shear modulus at temperature $T$, $G_o=3.76$  GPa is shear modulus for the dry FRP sample, $\alpha$ is model parameter and $t$ is time in days. An exponential trend is chosen for its simplicity, requiring only one model parameter $\alpha$. Furthermore, due to an exponential trend, shear modulus will tends to zero for degradation over a very long period, i.e. $\lim_{t\to\infty} G=0$. Comparison between the experimental and the exponential curves fitted to the experimental data (using the MATLAB Curve Fitting Toolbox) is shown in Figure \ref{Fig_GExp_Fit}, where values of the model parameter $\alpha$ are 0.0023, 0.0027 and 0.0040 for the exposure temperatures of $25^oC$, $60^oC$ and $80^oC$ respectively. Parameter $\alpha$ represents the rate of of degradation of the FRP material and increases with exposure temperature. In generalised form, for any exposure temperature $T$, Equation (\ref{Eq_ExpTrend}) is written as 
\begin{equation}\label{Eq_Model1}
G\left(T, t\right)=G_oe^{-\alpha\left(T\right) t},
\end{equation}
where $\alpha\left(T\right)$ is function of temperature and is written as 
\begin{equation}\label{Eq_alphaT}
\alpha\left(T\right)=\beta \textrm{ln}\left[1-\frac{T}{T_g}  \right],
\end{equation}
where $T_g$ is the glass transition temperature and $\beta$ is a model parameter. In this case, an experimentally calculated glass transition temperature of 126 $^oC$ is used. The specific choice of Equation (\ref{Eq_alphaT}) is based on the general observation that the degradation stops at $T=0K=-273.15 ^oC$ and increases rapidly at $T \rightarrow T_g$. Using the values of $\alpha$ already estimated for the exposure temperature of $25^oC$, $60^oC$ and $80^oC$, a value of $\beta$ can be estimated using least square fitting, i.e. minimisation of the following equation w.r.t. $\beta$
\begin{equation}\label{Eq_alphaT_LSF}
F\left(\beta\right)=\sum_{i=1}^{3}\left(\alpha_i - \beta \textrm{ln}   \left[1-\frac{T_i}{T_g}  \right]  \right)^2,
\end{equation}
leading to a value of $\beta=-0.001682$. Using this value, a comparison with the fitted exponential curves is shown in Figure \ref{Fig_GExp_Fit} indicating very good agreement. Furthermore, the least square fitted curve adheres to the aforementioned assumption for the rate of degradation at $T=0K$ and $T=T_g$. Combining Equation (\ref{Eq_Model1}) and (\ref{Eq_alphaT}), the degradation model is written as 
\begin{equation}\label{Eq_Model2}
G\left(T, t\right)=G_oe^{-\beta \textrm{ln}\left[1-\frac{T}{T_g}  \right] t}.
\end{equation}
A comparison between the experimental data, fitted exponential curves and the proposed degradation model (Equation (\ref{Eq_Model2})) is shown in Figure  \ref{Fig_GExp_Fit_Model}.
\begin{figure}[h!]
\begin{centering}
\includegraphics[scale=0.7]{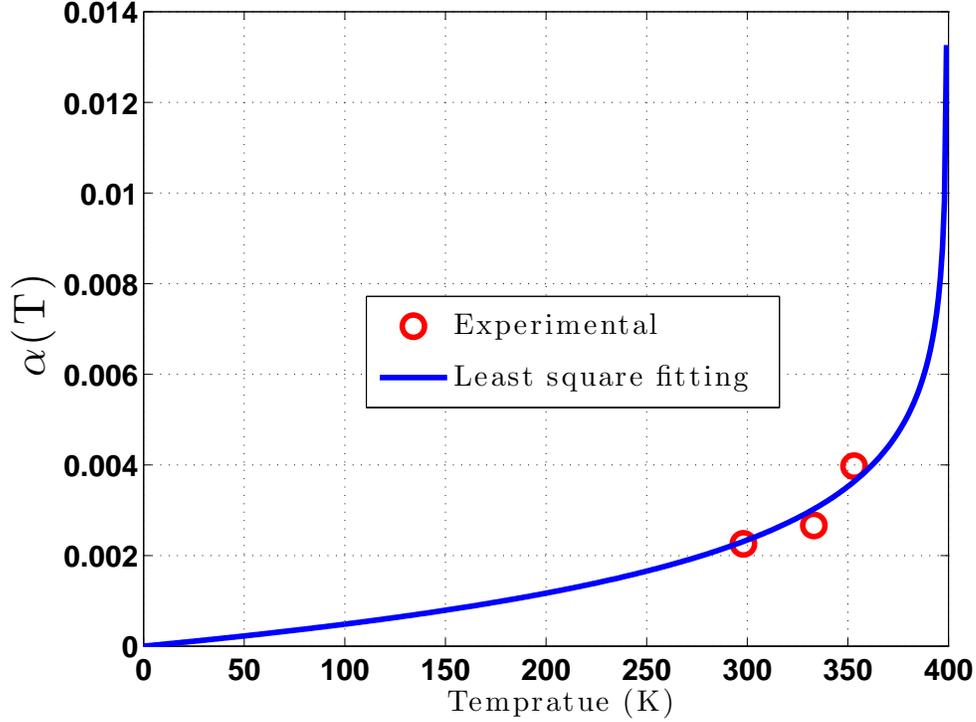}
\caption{$\alpha(T)$ versus temperature (experimental and least square fitting)} \label{Fig_Galpha}
\end{centering}
\end{figure}

\begin{figure}[h!]
\begin{centering}
\includegraphics[scale=0.9]{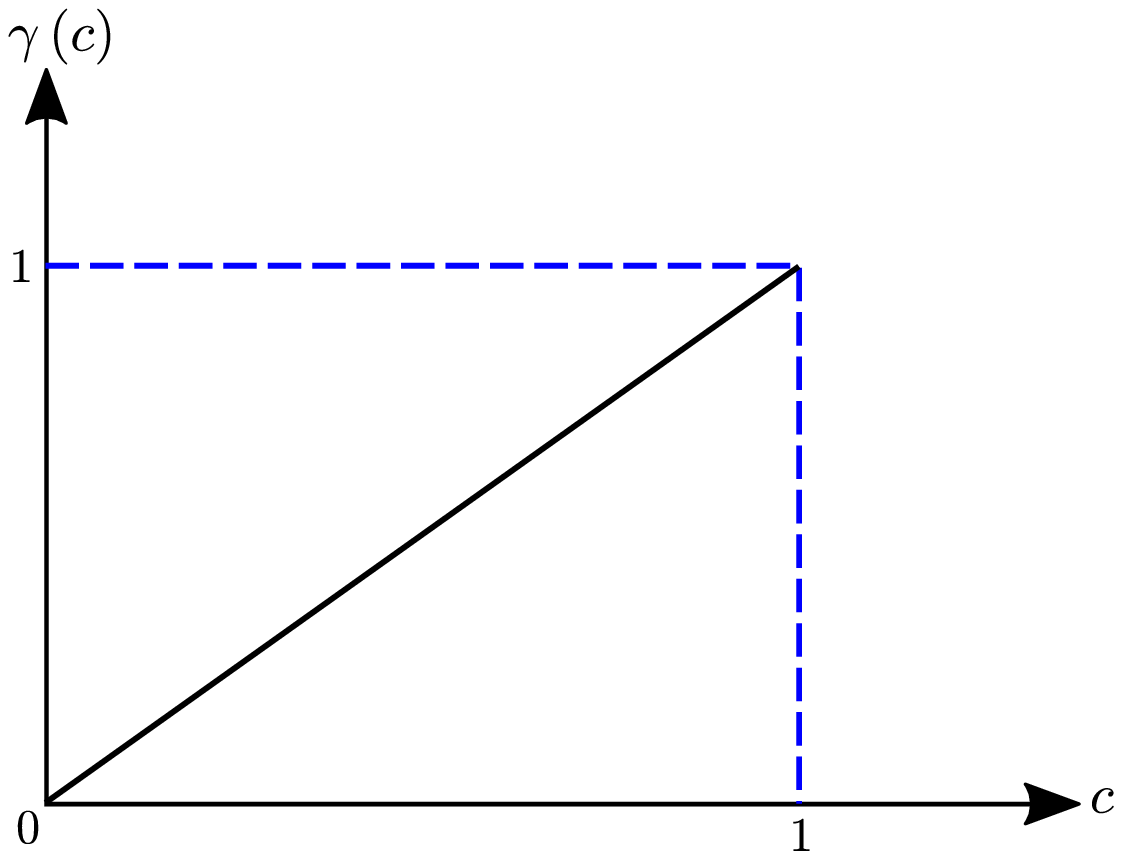}
\caption{$\gamma\left(c\right)$ versus moisture concentration $c$} \label{Fig_Gamma_c} 
\end{centering}
\end{figure}

The degradation model in Equation (\ref{Eq_Model1}) is next adapted to include the degree of moisture exposure. A new model parameter is introduced, $\gamma$, that is dependent on moisture concentration, $c$, and Equation (\ref{Eq_Model2}) is modified as follows:
\begin{equation}\label{Eq_Model3}
G\left(T, t\right)=G_oe^{-\beta \gamma\left(c\right) \textrm{ln}\left[1-\frac{T}{T_g}  \right] t}.
\end{equation}
In this paper, due to the lack of experimental data, a simple linear relationship for $\gamma\left(c\right)$ is proposed (shown in Figure \ref{Fig_Gamma_c}), i.e. for fully saturated sample (e.g. sample fully immersed in water) $\gamma\left(c=1\right)=1$ and  $\gamma\left(c=0\right)=0$. Thus Equation (\ref{Eq_Model3}) becomes
\begin{equation}\label{Eq_Model_final1}
G\left(T, c, t\right)=G_oe^{-c\beta \textrm{ln}\left[1-\frac{T}{T_g}  \right] t}.
\end{equation}
Equation (\ref{Eq_Model_final1}) is next modified for the real case of temporally varying temperature $T\left(t\right)$ and moisture concentration $c\left(t\right)$, which in the rate form is written as 
\begin{equation}
\frac{d}{dt}G\left(T,c,t\right)=\frac{\partial G}{\partial T}   \cancelto{0}{\frac{\partial T}{\partial t}}  +\frac{\partial G}{\partial c}  \cancelto{0}{\frac{\partial c}{\partial t}}+\frac{\partial G}{\partial t}=-c\beta\textrm{ln}\left(1-\frac{T}{T_{g}}\right)G.
 \end{equation}
We have assumed that the degradation process is very slow compared to the daily variation of temperature $T$ and moisture concentration $c$. This degradation model is now generalised for the matrix isotropic damage, with the assumption that it degraded in a similar way as shear modulus, and is expressed as
\begin{equation}\label{Eq_Model_RateForm}
\frac{d}{dt}\left(1-\omega \right)=-c\beta\textrm{ln}\left(1-\frac{T}{T_{g}}\right)\left( 1-\omega\right), 
\end{equation}
where $\omega$ is the isotropic damage parameter, with $\omega=0$ equivalent to no degradation and $\omega=1$ is full degradation. This degradation model is discretised in time using an implicit Backward Euler Method.

\section{Computational framework}\label{sec_comp_framework}
\begin{figure}[h!]
\begin{centering}
\includegraphics[scale=0.7]{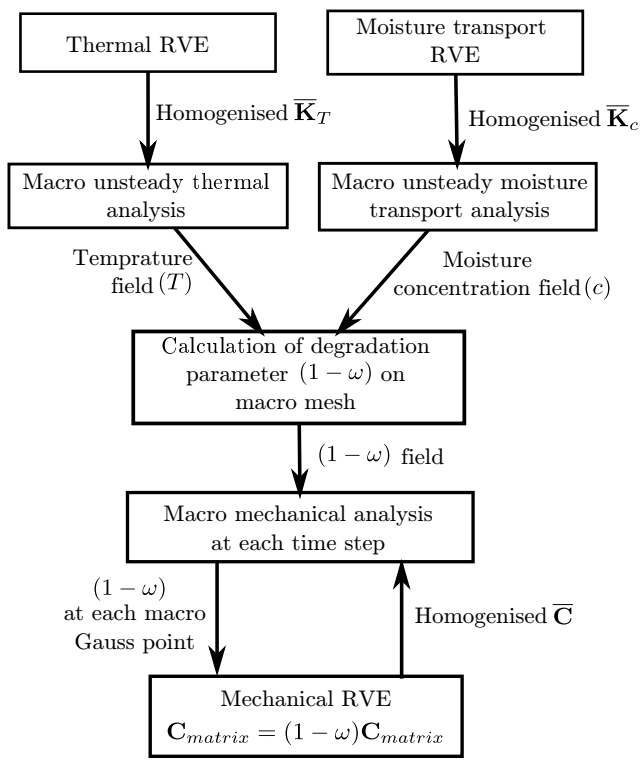}
\caption{Flow chart for the computational framework} \label{Fig_Comp_Framework} 
\end{centering}
\end{figure}

The proposed computational framework, composing a series of micro and macro-level analyses, is implemented in our group's FE software MoFEM (Mesh Oriented Finite Element Method) \cite{MoFE_2015}. The detailed flow chart for the computational framework is shown in Figure \ref{Fig_Comp_Framework}. In this paper, one-way coupling is assumed, i.e. mechanical analysis is dependent on thermal and moisture transport analyses but not vice-versa. Furthermore, it is also assumed that thermal and moisture transport analyses are independent of each other. The conductivity matrices $\overline{\mathbf{K}}_T$ and $\overline{\mathbf{K}}_c$ for these analyses are determined from CH of the underlying RVEs with known conductivities for matrix and yarns materials. Both $\overline{\mathbf{K}}_T$ and $\overline{\mathbf{K}}_c$ are calculated only once and are used subsequently for each Gauss point and every time step for the macro-level heat transfer and moisture transport analysis respectively. At each time step, temperature $T$ and moisture concentration $c$ fields are saved to the macro-mesh, which are used to calculate the degradation parameter $1-\omega$. 

The residual associated with Equation (\ref{Eq_Model_RateForm}) is written as
\begin{equation}\label{Eq_Deg_residual}
F= \frac{d}{dt}\left(1-\omega \right)+c\beta\textrm{ln}\left(1-\frac{T}{T_{g}}\right)\left( 1-\omega\right).
\end{equation}
Discretisation of the degradation parameter follows the standard form:
\begin{equation}\label{Eq_Deg_approx}
\left(1- \omega \right)^h =\boldsymbol{\mathcal{NW}}_{\left(1- \omega \right)},
\end{equation}
where $\left(1- \omega \right)^h$ is the approximation of $\left(1- \omega \right)$ and $\boldsymbol{\mathcal{W}}_{\left(1- \omega \right)}$ is vector of unknown nodal values of ${\left(1- \omega \right)}$. In the discretised form Equation (\ref{Eq_Deg_residual}) is written as 
\begin{equation}\label{Eq_Deg_residual_disc}
\mathbf{F}=\int_{\Omega}\boldsymbol{\mathcal{N}}^{T}\left(\frac{d\boldsymbol{\mathcal{NW}}_{\left(1-\omega\right)}}{dt}+c\beta\textrm{ln}\left(1-\frac{\boldsymbol{\mathcal{N}}\mathbf{T}}{T_{g}}\right) \boldsymbol{\mathcal{NW}}_{\left(1-\omega\right)}\right)d\Omega
\end{equation}
Furthermore, the Jacobean associated with Equation (\ref{Eq_Deg_residual}) is written as 
\begin{equation}\label{Eq_Deg_jac}
J=\frac{\partial F}{\partial\left(1-\omega\right)}+a\frac{\partial F}{\partial\overset{\centerdot}{\left(1-\omega\right)}},
\end{equation}
where $\overset{\centerdot}{\left(1-\omega\right)}= \frac{d\left(1-\omega\right)}{dt}$ and $a$ is a positive shift, the value of which depends on the integration scheme used. In discretised form Equation (\ref{Eq_Deg_jac}) is written as 
\begin{equation}\label{Eq_Deg_jac_disc}
\mathbf{J}=\int_{\Omega}\boldsymbol{\mathcal{N}}^{T}\left(c\beta\textrm{ln}\left(1-\frac{\boldsymbol{\mathcal{N}}\mathbf{T}}{T_{g}}\right)+a\right)d\Omega
\end{equation}
Solution of Equations (\ref{Eq_Deg_residual_disc}) and (\ref{Eq_Deg_jac_disc}) leads to field $\left(1-\omega\right)$ on the macro-mesh for each time step. At each time step, the macro-mechanical problem consisting of linear elastic analysis is next solved using the fully nested FE$^2$ algorithm. At each macro-level Gauss point, the degradation parameter $\left(1-\omega\right)$ is calculated and transferred to the underlying mechanical RVE. Degradation is considered only for the matrix part of the mechanical RVE; the fibres remain undegraded at present. The homogenised matrix $\overline{\mathbf{C}}$ is calculated for each mechanical RVE and passed back to the macro-Gauss point to be used in the subsequent macro-mechanical analysis.

\section{Heuristic method for computation of yarn directions}\label{sec_fibre_directions}
In order to re-orient the known stiffness matrix for the transversely isotropic material from the local coordinates to global coordinates, the yarn directions at each Gauss point need to be determined. To do this it is possible to simply use the cubic splines that were used to construct the yarns. However, this can lead to inaccuracies in the case of yarns with non-uniform cross-sections along their length. An alternative approach is proposed here in which the yarn directions are determined by solving the potential flow field separately along each yarn. The governing equation for potential flow is given as
\begin{equation}   
\nabla^{2}\phi=\frac{\partial^{2}\phi}{\partial {y_1}^{2}}+\frac{\partial^{2}\phi}{\partial {y_2}^{2}}+\frac{\partial^{2}\phi}{\partial {y_3}^{2}}=0, 
\end{equation}
where $\phi$ is the stream function, the gradient of which, i.e. $\nabla\phi=\mathbf{v}$ subsequently determines the yarn directions. A detailed description of how to transfer the stiffness matrix for transversely isotropic material between local and global coordinate axes is given in previous publications \cite{Ullah2014, Michael2014, Xiaoyi2015}. 

\section{Numerical example}\label{sec_num_exp}
\begin{figure}[h!]
\begin{centering}
\includegraphics[scale=0.7]{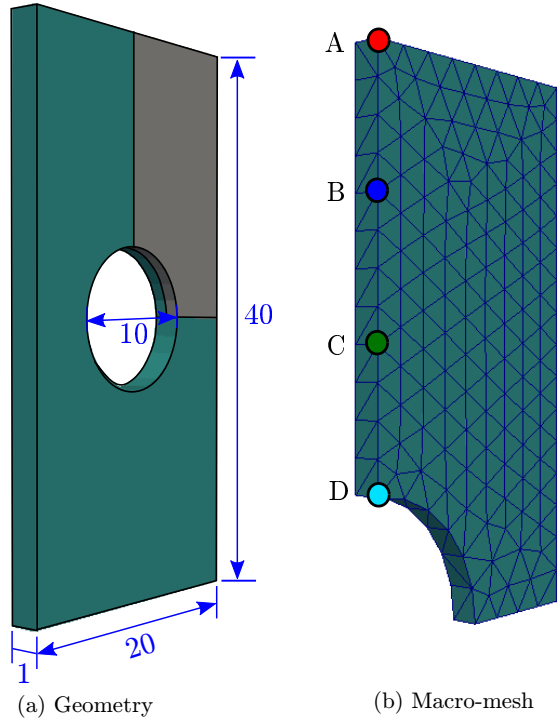}
\caption{Macro-level geometry and mesh for the numerical example} \label{Fig_NE_GeomMesh} 
\end{centering}
\end{figure}

A macro-level three-dimensional plate structure with a hole is considered, the geometry for which is shown in Figure \ref{Fig_NE_GeomMesh}(a).  For the macro-level thermal problem a temperature of 80 $^oC$ is applied to the top and bottom surfaces and constant heat flux is applied to the left and right surfaces. Similarly, for the moisture transport problem, a constant concentration of 1 (100\% RH) is applied to the top and bottom surfaces and constant moisture flux is applied to the left and right surfaces. For the macro-mechanical analysis, a uniform pressure of 1000 MPa is applied along the upper and lower faces in the vertical direction. Due to symmetry of the geometry and boundary conditions for heat transfer, moisture transport and mechanical analysis, only one-eighth of the problem, shown in grey in Figure \ref{Fig_NE_GeomMesh}(a) is considered. 

\begin{figure}[h!]
\begin{centering}
\includegraphics[scale=0.7]{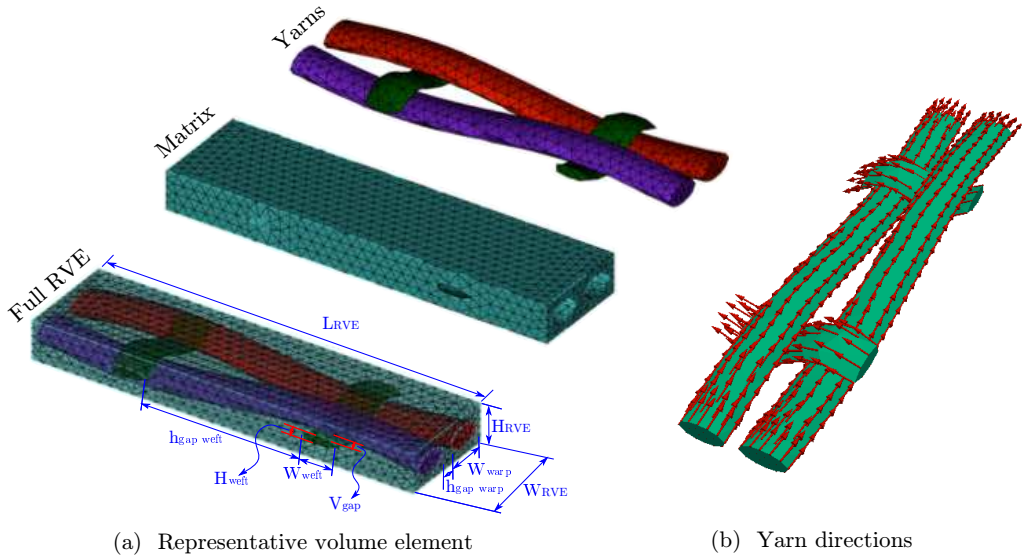}
\caption{Representative volume element geometry and yarn directions} \label{Fig_NE_RVE_Geom} 
\end{centering}
\end{figure}

The RVE used in this case, comprising a plane wave textile composite, is shown in Figure \ref{Fig_NE_RVE_Geom}(a). The geometry parameters for the RVE are shown in Table \ref{Table_NE_Table_RVE_Geom}. Both macro-structure and RVE are discretised with tetrahedral elements with 780 elements in the case of macro-structure and 10,285 elements in the case of the RVE.  An example of the yarn directions for the currently used RVE calculated from the potential flow analysis is shown in Figure \ref{Fig_NE_RVE_Geom}(b). For the matrix material, thermal conductivity, density, specific heat and moisture diffusivity are 190 kg mm/s$^3$C, 1.2x10$^{-6}$ kg/mm$^3$, 805x10$^6$ mm$^2$/s$^2$C and 2.8�10$^{-6}$mm$^2$/s respectively, while the corresponding values used for the yarns, with the same units, are 1030, 2.53x10$^{-6}$, 1000x10$^6$, 1.46�10$^{-7}$, 230 and 0.26. For the mechanical RVE analysis, the matrix and yarns are assumed to be isotropic and transversely isotropic materials respectively. The material properties used are shown in Table \ref{Table_NE_Table_RVE_mat}.

\begin{figure}[h!]
\begin{centering}
\includegraphics[scale=0.7]{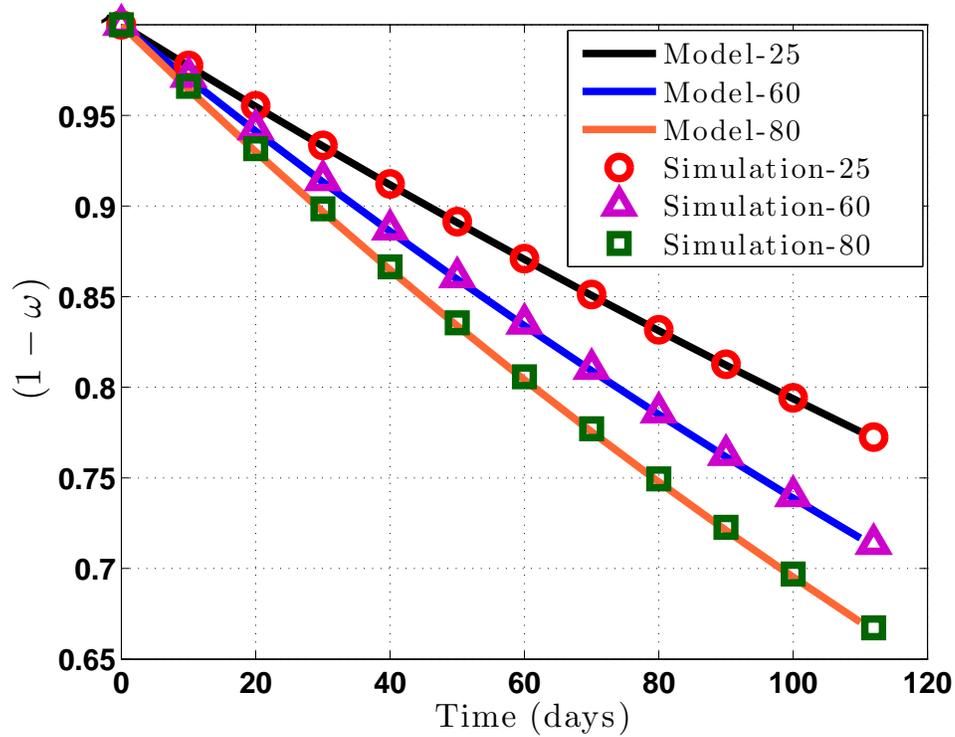}
\caption{$\left(1-\omega\right)$, simulation versus model results } \label{Fig_NE_WtModelVsSimulation} 
\end{centering}
\end{figure}

The correct numerical implementation of the degradation model on the macro-mesh, shown in Figure \ref{Fig_NE_GeomMesh}(b), is verified by exposing it to temperatures of 25$^oC$, 60$^oC$ and 80$^oC$ for 112 days. The moisture concentration is kept at 100\% throughout the domain. Comparison between the simulation results recorded at point B (shown in Figure \ref{Fig_NE_GeomMesh}(b)) and the model prediction using the degradation model given in Equation (\ref{Eq_Model_final1}) is shown in Figure \ref{Fig_NE_WtModelVsSimulation}. Both degradation model and simulation results are in excellent agreement, verifying the numerical implementation. 

\begin{figure}[h!]
\begin{centering}
\includegraphics[scale=0.7]{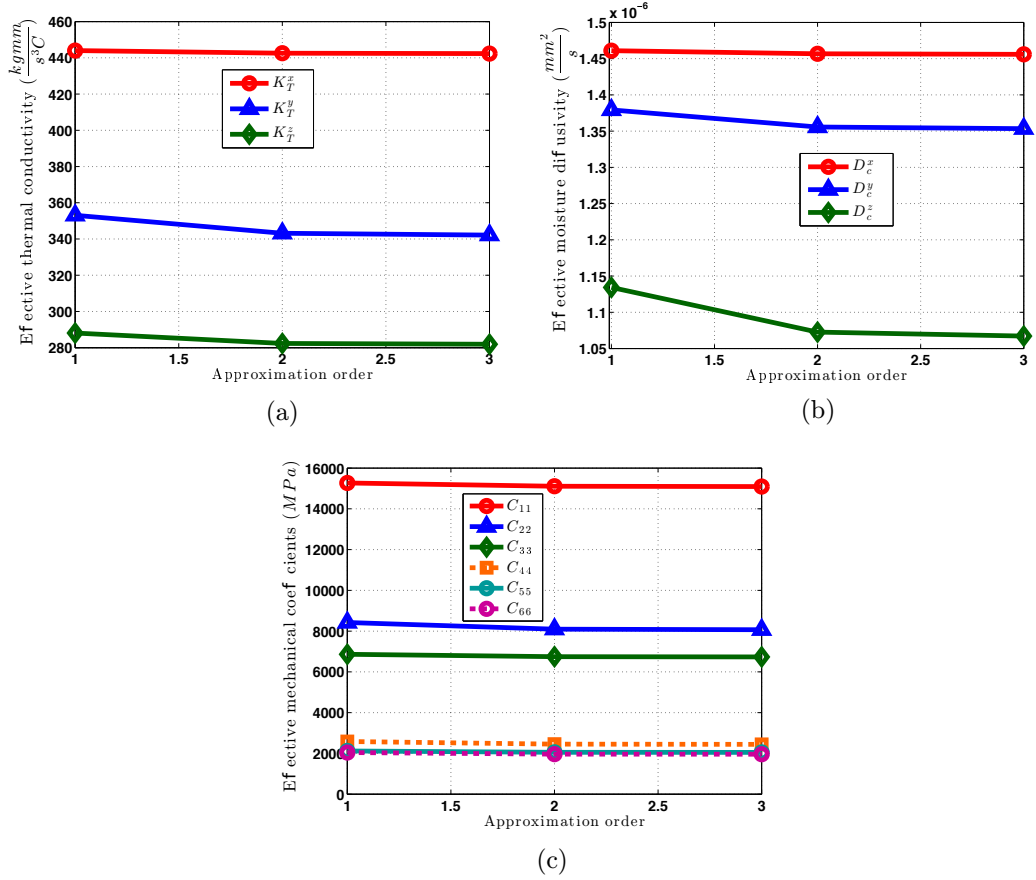}
\caption{Diagonal elements of the homogenised matrices $\overline{\mathbf{K}}_T$, $\overline{\mathbf{D}}_c$ and $\overline{\mathbf{C}}$ in the case of heat transfer, moisture transport and mechanical RVEs} \label{Fig_Coeff_Conversion} 
\end{centering}
\end{figure}

Before starting the actual multi-scale and multi-physics analysis, convergence studies are performed for all the three type of RVEs (heat transfer, moisture transport and mechanical) for the calculation of the homogenised matrices. In the convergence analysis, $1^{st}$, $2^{nd}$ and $3^{rd}$ order approximations are used  based on hierarchic basis functions \cite{Ainsworth2003}. The corresponding degrees of freedom in the case of heat transfer and moisture transport are 2,364, 16,214 and 51,836 respectively, while for mechanical case the degrees of freedom are 7,092, 48,642 and 155,508 respectively. Diagonal elements of the homogenised matrices $\overline{\mathbf{K}}_T$ and $\overline{\mathbf{D}}_c$  and $\overline{\mathbf{C}}$ in the case of heat transfer, moisture transport and mechanical RVEs are plotted versus the order of approximation and are shown in Figure \ref{Fig_Coeff_Conversion}(a-c). There are some noticeable difference as a result of increasing the order of approximation from 1 to 2 but further increasing the orders of approximation has no significant effect. Therefore, approximation order of 2 is selected for the CH for all three types of RVEs, as the computational cost will increase dramatically with $3^{rd}$ approximation with very little effect on the results.

\begin{figure}[h!]
\begin{centering}
\includegraphics[scale=0.7]{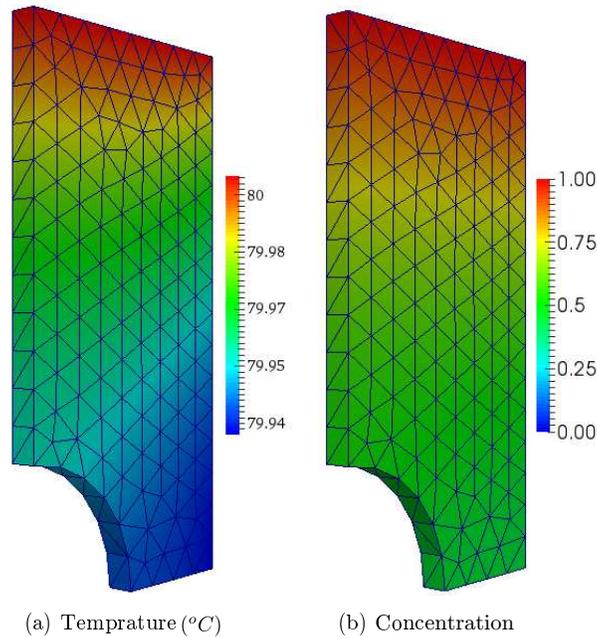}
\caption{Temperature and moisture concentration fields at the end of simulation for the numerical example} \label{Fig_NE_TempConc} 
\end{centering}
\end{figure}

\begin{figure}[h!]
\begin{centering}
\includegraphics[scale=0.7]{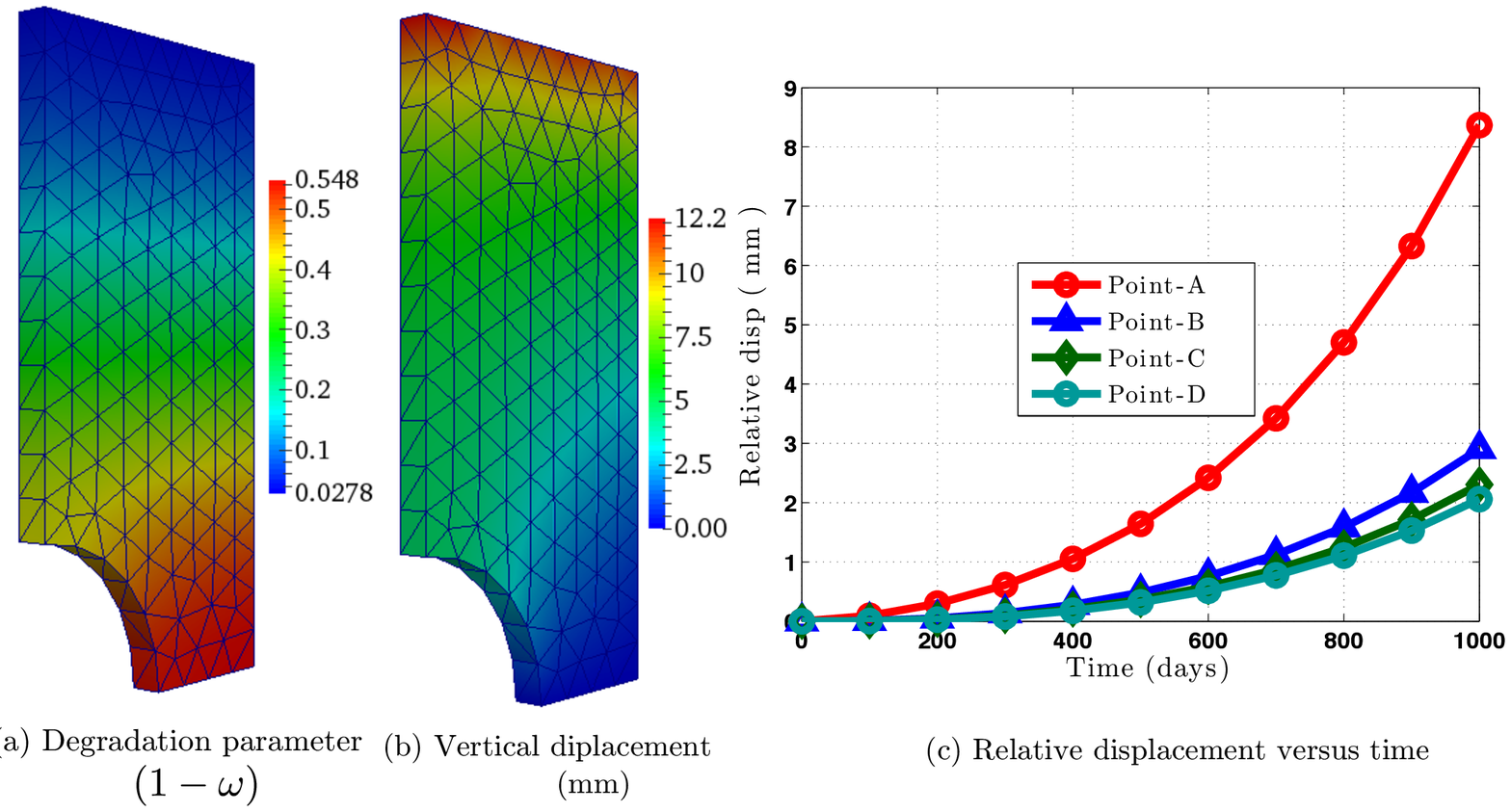}
\caption{$\left(1-\omega\right)$, vertical displacement fields and relative displacement versus time for the numerical example} \label{Fig_NE_WtUy} 
\end{centering}
\end{figure}

Both macro-level thermal and moisture transport problems are solved with time step of 10 days for total time of 1000 days. Heat transfer reaches equilibrium very quickly as compared to the moisture transport analysis due to its higher value of thermal conductivity. At the end of the simulation, temperature and moisture concentration fields are shown in Figures \ref{Fig_NE_TempConc}(a) and \ref{Fig_NE_TempConc}(b) respectively. Furthermore, at each time step the damage variable $\left(1-\omega\right)$ is calculated from both the temperature and moisture concentration fields, see Figure \ref{Fig_NE_WtUy}(a). As expected, the area near the top surface of the plate degraded more as compared to the bottom surface due to the higher temperate and moisture concentration. Vertical displacement $u_y$ at the end of simulation is shown in Figure \ref{Fig_NE_WtUy}(b), while vertical displacements relative to the first time step for points A, B, C and D (identified in Figure \ref{Fig_NE_GeomMesh}(b)) are shown in Figure \ref{Fig_NE_WtUy}(c). It is clear that this increase in vertical displacement at constant load is due to the degradation of the macro-structure stiffness.

\begin{table}[h!]
\caption {RVE geometry parameters (all dimension in mm)} \label{Table_NE_Table_RVE_Geom} 
\centering
\begin{tabular}{ll|ll}
\hline
Parameters & Values & Parameters & Values \\
\hline
W$_{warp}$ & 0.3 & W$_{weft} $ & 0.3 \\
H$_{warp}$  & 0.1514 & H$_{weft} $ & 0.0757 \\
h $_{gap \,warp} $ & 0.09 & h$_{gap \,weft} $ & 1.2 \\
L$_{RVE }$ & 3.0 & v$_{gap} $ & 0.012 \\
W$_{RVE} $ & 0.78 &  &  \\
H$_{RVE} $ & 0.3 &  &  \\
\end{tabular}
\end{table}

\begin{table}[h!]
\caption {RVE mechanical parameters (all modulii in GPa)} \label{Table_NE_Table_RVE_mat} 
\centering
\begin{tabular}{lllll|ll}
\hline
\multicolumn{5}{c|}{Fibres properties} & \multicolumn{2}{c}{Matrix Properties} \\
\hline
$E_p$ & $E_z$ & $\nu_p$ & $\nu_z$ & $G_{pz}$ & $E$ & $\nu$ \\
17.5 & 35 & 0.26 & 0.26 & 8.75 & 3.5 & 0.3 \\
\end{tabular}
\end{table}

\section{Concluding remarks}\label{sec_conc_remarks}
In this paper, an ongoing work, consisting of a coupled hygro-thermo-mechanical computational framework based on multi-scale CH is described for textile based FRP composites. A degradation model is developed from the experimental data for the mechanical properties of FRP composites and is incorporated in the computational framework. The numerical formulation and corresponding implementation of the degradation model are described. Numerical implementation of the RVE unified boundary conditions and the associated computation of homogenised stresses and stiffness matrix are also given in detail. For the accurate modelling of the textile RVE for the thermal, moisture transport and mechanical analyses, convergence studies based on the hierarchic basis functions are also conducted.  For the mechanical analysis, matrix and yarns are considered as isotropic and transversely isotropic materials respectively. For the transversely isotropic materials, principal directions are calculated using potential flow analysis. The proposed computational framework is designed to take advantage of high performance computing for distributed memory computer architectures, leading to an efficient code even for large computations associated with multi-scale and multi-physics algorithms. A three-dimensional numerical example is presented to demonstrate the correct implementation and performance of the proposed computational framework. The degradation model is sufficiently general to allow it to be applied to different composites and fitted to different experimental data.    

\section*{Acknowledgements}
The authors gratefully acknowledge the support of the UK Engineering and Physical Sciences Research Council through the Providing Confidence in Durable Composites (DURACOMP) project (Grant Ref.: EP/K026925/1).

\bibliographystyle{elsarticle-num}
\bibliography{bibfile1}

\end{document}